# A collective insight into the cultural and academic journeys of Native Hawaiians while pursuing careers in physics and astronomy

Input for the Astro2020 Decadal Topic:
The State of the Profession and Social Impacts (SoP)

*Authors (in alphabetical order):*

Heather Kaluna
Department of Physics & Astronomy
University of Hawai'i at Hilo
kaluna@hawaii.edu

Mailani Neal
Department of Physics
New Mexico Institute of Mining and Technology
mailanineal@gmail.com

Makana Silva
Department of Physics
The Ohio State University
silva.179@osu.edu

Tyler Trent
Department of Physics
University of Arizona
ttrent@email.arizona.edu

## Introduction

In the field of astronomy, Maunakea is known as a prestigious site for observing and science. In Native Hawaiian culture, Maunakea is revered as the connection between past, present, and future generations and their ancestral lands of Hawaiʻi. We have reached a juncture at which it is necessary to allow and enable Native Hawaiians to pursue careers in astronomy, especially on Maunakea. This paper serves to tell the accounts of four Kanaka astronomers and raise awareness of the barriers they have faced while pursuing astronomy careers. The authors identify issues that the community faces due to the disconnect between astronomy and Hawaiʻi communities and propose resolutions to lead the way forward.

## Heather Kaluna

I was born, raised and educated in the rural town of Pāhoa, located near the eastern most part of Hawaiʻi Island. Like many other indigenous communities, Pāhoa is one of the lowest income regions within the Hawaiian islands. However, Pāhoa's richness lies within its fishing and farming traditions, as well as the tight-knit relationships that result from the Hawaiian perspectives of family and community. *Nonetheless, while the continual support and encouragement of my family and community allowed me to succeed in school, it could not make up for the financial challenges that I faced when entering into college as a first generation college student.* After my first year of college, the combined lack of my college readiness and my parents' lack of experience in the college financial aid process lead to my dropping out of school as I no longer had funding to pay for it. Were it not for a strange twist of fate and a timely increase in my father's income, which came after decades of trying to verify his disability as a product of his Vietnam service, I likely would not be where I am today. Fortunately, after returning to school and working hard to not make a similar mistake, I was able to eventually graduate and pursue a PhD in astronomy.

I now serve as an educator and researcher at my alma mater, the University of Hawaiʻi at Hilo (UH Hilo). Although I grew up only vaguely being aware of the research activities taking place on Maunakea, the development of my career in astronomy is intimately tied to the relationship, both spiritual and academic, that I have built with this mountain. Oddly enough, although I only lived a few hours away from the mountain, my first visit to Maunakea was not until my first year of college. The visit was prompted by a class field trip to tour the observatories, and thus it was the modern practice of astronomy that led me to better understand my sense of place. Although my courses peaked my interest and led to my pursuit of a bachelor's degree that would prepare me for an astronomy career, my first real introduction to astronomical research came from a summer internship through the Akamai Internship Program. The aim of this program is to promote the advancement of local students within STEM careers here in Hawaiʻi. Many of the interns are placed with the observatories based here in Hawaiʻi, however my own experience took me to the University of California, Santa Cruz. Although I was not working directly with an observatory here in Hawaiʻi, my research project involved the analysis of data collected using the Keck observatories on Maunakea. *This internship provided me with my first authentic experience in connecting with the research culture of Maunakea.* It was this uniquely positive research experience that really inspired me to pursue a PhD in the field of astronomy.

It is with much regret that during my graduate school experience, I was unable to supplement my academic training with a dual training in Native Hawaiian cultural practices. More often than not, *my cultural perspectives are treated as irrelevant or insignificant.* However, through initiatives such as Hawaiʻi Papa o Ke Ao and Uluakea here at UH Hilo, I have been encouraged to incorporate my cultural perspectives into the instructional and academic experiences that I design for my students. I have also been *given opportunities to expand upon my knowledge of Hawaiian cultural practices, and use these to improve the educational experience of my students.*

*The challenges I have faced on my path to a career in astronomy are all too frequent among our local and Native Hawaiian students.* Thus, I believe it is important to highlight these barriers within this document so that we may better address the issues that our Hawaiʻi students face and increase the representation of our local and native students within astronomy.

**Makana Silva**

Aloha mai kākou, ʻo Makana Silva koʻu inoa. No Mākāhā mai au aka noho au i Columbus, Ohio no ke kiʻi ʻana o kaʻu laeʻula kālaikūlohea. He Hawaiʻi au. Aloha everyone, my name is Makana Silva. I am from Mākāhā, Oʻahu but I live in Columbus, Ohio working to obtain a PhD in astrophysics. I am Hawaiian. As a Hawaiian, I stand behind and support projects that come to Hawaiʻi that promote education and make an effort to build connections with the surrounding communities, such as the Thirty Meter Telescope (TMT).

Growing up on the West side of Oʻahu, I had virtually no exposure to the idea of science, technology, engineering, and math (STEM) related careers. The public school system that I was in just didn't have the resources to publicize and promote to middle and high school kids the opportunities in STEM. It wasn't until I attended Kamehameha Schools, a private school, in my sophomore year of high school did I receive the necessary exposure and immersion of STEM related pathways that I knew that I was gonna pursue physics as my lifetime journey. But even then, I was still unaware as to how physics was integrated within my community, my home. It wasn't until the beginning of graduate school that I learned just how special and pivotal my home is for cutting edge research in astronomy and astrophysics. I had to leave Hawaiʻi just to understand the magnitude with which my home played in the field that I want to contribute to. *This is an example of the disconnect between the communities of Hawaiʻi and the institutions conducting research within.* I believe one way to build a stronger connection is to promote and support STEM opportunities and research at all levels of education in the community. For myself and most kids in Hawaiʻi, it was only through external scholarships that allowed me to pursue higher education and ultimately get to where I am today. In the case of TMT, they are providing financial support to education in the form of the THINK Fund as well as promoting STEM opportunities through the Workforce Pipeline Program. For me, these funds really brought to life my dreams of being a researcher and coming back to Hawaiʻi and sharing what I learned with my community.

       In July of 2019, I was in Amsterdam where I had the awesome opportunity to attend the Advancing Theoretical Astrophysics Summer School. It was during this time I saw on social media the arrests of protestors commencing on the access road to the summit of Maunakea. Regardless of my personal support of the TMT, it was painful to see. To see members of my own ʻohana arresting and getting arrested was heartbreaking and that is when I had to really dig deep into my "naʻau" (gut feeling) and ask for guidance. I had to ask if my support for this project really just hurting my community more than helping it. That's what I live for, to come home and help my people and help improve the quality of life in Hawaiʻi. I spent about a week just staying to myself and exploring the realms of my beliefs and my identity as a Hawaiian. And then it hit me, I don't support TMT because I love physics, I support TMT because I am Hawaiian. Our people have a rich history of being multifaceted and excelled in art, science, technology, medicine, etc. Any ancient indigenous culture is remembered by what they leave behind, their legacy. For Hawaiians, our legacy includes sailing across thousands of miles of open sea only by the knowledge of the stars and movement of the oceans. The very same stars we are studying today were the ones my ancestors looked upon for guidance in the darkness in the open waters. My people have always and still study and learn about the world around us, which is exactly what TMT as a tool is going to accomplish. This is why I support TMT, it is an opportunity for Hawaiians to add to this legacy, *perpetuation through contribution*. I believe the mission of TMT is a modern approach to practicing Hawaiian culture and an opportunity for Hawaiians to become leaders in this global endeavor of scientific discovery. It is time to support a project that comes to Hawaiʻi that will give Hawaiians advancement in education and create a more diverse job market other than tourism. *TMT is an example of the sort of projects we need to encourage in Hawaiʻi because it is a tool that we can use as a way to practice Hawaiian culture in a modern way and integrate this culture within the diverse field of astronomy and astrophysics.*

**Mailani Neal**
I am a Native Hawaiian who was born and raised on Hawaiʻi Island. I am currently pursuing a PhD in Physics and Astronomical Instrumentation at New Mexico Institute of Mining and Technology. I hold a bachelor's degree in Applied Physics with a concentration in astronomical instrumentation from Rensselaer Polytechnic Institute. I hope my career will create the undoubtedly sensible realization that we, the indiginous people of Hawaiʻi, are highly capable of contributing to the forefront of innovation, especially that which occurs on our ʻāina (land).

When I was nine years old, my science and social studies teacher introduced our class to a unit about Hawaiian voyaging and navigation practices. She taught this in conjunction with a science unit about astronomy. I was absolutely amazed that my ancestors were so knowledgeable of the night sky and used that knowledge to become the greatest voyagers of all time. I felt that astronomy could be the bridge between two aspects of myself: being a Native Hawaiian and being a modern day scientist. I continue to believe this. I found my cultural identity as a Native Hawaiian because of astronomy and my life calling to be an astronomer through my responsibility to perpetuate the Hawaiian culture. Unfortunately, the elusiveness of astronomy and many professional science fields makes that statement sound and feel contradicting.

Since my experience at nine years old, my experiences in astronomy and Hawaiian Culture have been segregated. I was fortunate to attend a private school for Native Hawaiian children. I studied ʻŌlelo Hawaiʻi (Hawaiian Language) for five years and had the opportunity to learn about my Hawaiian culture through days volunteering in the loʻi (irrigated terraces for taro) and loko iʻa (fish pond) along with other experiences.

I attended (the now defunct) HI-STARS summer camp at the University of Hawaiʻi Mānoa. I spent the summers of 2017 and 2018 as an intern at the East Asian Observatory in Hilo, Hawaiʻi. This was a dream come true: *I am a Native Hawaiian astronomer of Maunakea*.

Opportunities like this are, sadly, not easily accessible to the local community for a variety of reasons. I was able to have these opportunities because of my supportive parents, education mentors, and generous individuals in the Big Island astronomy community. One organization that has enabled me to find these opportunities is The Thirty Meter Telescope International Observatory. TMT granted me scholarships through The Hawaiʻi Island New Knowledge (THINK) Fund so that I could afford college. This enabled me to attend a prestigious college that has greatly boosted my chances when I applied for internships, jobs, and graduate school. The connections that TMT helped me to establish with the astronomy community on Big Island led to the internships I had at the East Asian Observatory. TMT has also connected and provided me with the opportunity to be involved with my community through an internship with the Hawaiʻi Island Economic Development Board to encourage local students to pursue education further than a high school diploma. I am eternally grateful for those opportunities, therefore, I want to ensure that even more students can be afforded the same resources that I had while growing up.

This decadal survey is essential in determining the outcome of American astronomy and astrophysics over the next 10 years. *That is why I strongly urge the consideration of increasing opportunities and opening pathways into astronomy for Hawaiʻi students*. Historically, astronomy and astrophysics have always benefited from the involvement of many eyes on the sky. From the determination that the solar system is heliocentric, to the Shapley–Curtis Debate, the inclusion of women, such as Vera Rubin whose work contributes very greatly to our speculation of dark matter, and the collaboration of the Event Horizon Telescope, *astronomy and astrophysics have proven throughout history to be a field in which inclusion is a necessity.* Inclusion of Hawaiʻi students and the community that astronomy is a part of needs to be made a forefront of priority as our field traverses into the next decade.

**Tyler Trent**

I was born and raised on the island of Oʻahu. I am part Native Hawaiian, as well as Chinese, German, Filipino and Spanish. The main cultural influences on me growing up were Hawaiian and Chinese. To sum up how these two cultures influenced me, to celebrate New Years, my family would eat traditional Hawaiian food that you would find at any luau, and we would also pop firecrackers. At midnight, we would light a long string of firecrackers that would pop for over a minute just like how my great grandma would when she was living in China. Blending cultures is just how I was raised to be.

When I was a Boy Scout, to obtain the rank of Eagle Scout (the highest rank attainable in the Boy Scouts of America program), I had to design, organize, and lead an Eagle Scout Service Project. For my Eagle Scout Service Project, I chose to help restore an ancient Hawaiian fishpond by removing invasive seaweed in the spring of 2012. Fishponds were essential to the old Hawaiian way of life. By helping to restore this fishpond I was reconnecting with the ways of my ancestors, while also developing leadership skills as a Boy Scout. This is how I was being a Boy Scout in a Hawaiian cultural lens. Now I am a graduate student at the University of Arizona studying astrophysics.

My dream is to blend Hawaiian culture and modern astronomy. My path into astrophysics didn't start until I was in college. I was fortunate enough to come from a good private school that gave me enough of a foundation in math that pursuing a degree in STEM didn't seem impossible. However, my younger brother attended public school instead. When he was entering high school, he was unable to take the math class subsequent to the math he took in middle school because there were not enough seats available in the math class he needed. From that point on he lost interest in math. It was hard to motivate him to excel when he was limited by what the public school system had to offer. *I recommend more programs be created to help students coming from public schools, catch up and build their confidence to even dream of pursuing astronomy.*

For most of my high school and college years I came from a single parent home. With two brothers, one older and one younger, as soon as I was able to work, it was expected that I would contribute to the household. I worked at restaurants all throughout college so I could help pay for rent and groceries. It is only because I received multiple scholarships that covered my tuition, did it make attending the University of Hawaiʻi possible. *I recommend that more resources be put*

*into financial support for local students wanting to pursue degrees in astronomy and these resources be advertised.* The TMT's THINK fund is the type of program helping to provide scholarships for students pursuing STEM.

From a young age I learned about the Hawaiian culture. The private school I attended was a Hawaiian school that taught me many Hawaiian stories. Some even about Maunakea. However, *it wasn't until my junior year of college did I learn about the level of research being done on Maunakea.* When I learned about the research being done on the mountain, it made me proud that my home could contribute to the world's knowledge. It felt like a very Hawaiian thing for me to study the stars. My ancestors came to Hawaiʻi using their knowledge of the stars and now my home could be the premier location in the world to continue to learn about the celestial bodies above. However, if I didn't know about the level of research being done and saw how it comes from the same sense of wonder that native Hawaiians of old had, I would see the telescopes on the summit of Maunakea as inappropriately placed. *It is my recommendation that more resources be placed into outreach to share the amazing work being done on the mountain to the local communities.* This I believe will not only give a greater appreciation for the telescopes but also inspire the younger generations to study astronomy.

To blend two cultures it takes more than just the mutual acknowledgment of each other. There has to be more interdisciplinary research being done between Hawaiian culture and astronomy. An example of this is the work being done by Doug Simmons and Larry Kimura, on the physics of Pō. *I recommend supporting more work and programs (e.g. Physics of Pō, ʻImiloa) that work to bridge Hawaiian culture and modern astronomy.*

## Issues and Resolutions

We, the authors of this paper, have grown up in different families, on different islands, and attended different schools, yet in our experiences, we have faced similar issues. We summarize these recurring issues here and also make recommendations to further strengthen the relationships between the professional astronomy and local communities of Hawaiʻi.

**Economic Impediment**
Financial resources have been identified as one of the major barriers to the success and persistence of Native Hawaiian and other underserved students at the University of Hawaiʻi system[1].
In Hawaiʻi, the Thirty Meter Telescope has led the way in providing financial support for local students and benefiting the community. TMT is committing $1,000,000 to the lease rent of

which $800,000 will go towards the stewardship of Maunakea. No other telescope currently on the summit contributes this sort of financial effort just towards maintaining the integrity of the ʻāina (land). The THINK Fund allocated $1 million per year towards scholarships and STEM programs for students and teachers of Hawaiʻi Island.

- *Recommendation:* Support for programs that aim to address issues such as college readiness and financial support for our indigenous and underserved communities are essential to removing these economic barriers
- *Recommendation:* Promote projects that are invested in the communities in which they reside and projects that both aim to advance the field as well as the communities they serve

**Persistence and Advancement of Local/Indigenous Students in STEM**
Ensuring that have positive mentoring opportunities through summer internships and are able to identify themselves as scientists promotes the persistence of underrepresented students within STEM[2]. There are several programs that provide these opportunities in our Hawaiʻi community, however, a lack of consistent funding for these STEM mentoring programs can prove to be a barrier to their efficacy and success[2].

The Akamai Internship Program (AIP) has been successful in providing these opportunities to our local college students and highlighting the relevance of observatory-related careers[3]. However, AIP in particular is funded by the organizations in which the interns are placed, and as such, Akamai is especially vulnerable to fluctuations in funding that these organizations experience (e.g. the loss of TIO funding has drastically reduced the number of students in which this program can serve). Other programs such as HI-STAR[4], Maunakea Scholars and Journey through the Universe also have been successful in creating these interfaces between Hawaiʻi high school students and the professional astronomy community.

- *Recommendation:* Consistent and long-term support for the development and/or maintaining of these mentoring programs should be emphasized.

**Limited Interface between Hawaiian Cultural Perspectives and Astronomy in Academia**
Little emphasis has been given to address the lack of intersection between modern science and indigenous cultures[2]. We believe this can be achieved by further promoting traditional Hawaiian perspectives within the academic environment and research communities.

Uluakea[5] is a faculty development program that provides UH Hilo educators with the knowledge and experiences to design their curriculum from a place-based and culturally grounded approach.

By transforming the classroom experience into one of a culturally contexted place of learning, these programs can foster the intersections between modern science and cultural perspectives. The A Hua He Inoa[6,7] further emphasizes these intersections by enabling students to participate in modern astronomical research through a cultural lens.

- *Recommendation:* Support culturally-based professional development programs that aim to transform the classroom experience into a culturally contexted place of learning.
- *Recommendation:* Support organizations that foster educational opportunities through cultural immersion within a scientific framework of astronomical research